\documentclass[pra,aps,twocolumn,amsmath,amssymb,superscriptaddress]{revtex4-1}
\usepackage{graphicx}
\usepackage{amsfonts}
\usepackage{color}
\usepackage{amstext}
\usepackage{color}

\usepackage{epsfig,graphicx}
\usepackage{flafter}
\usepackage{amsmath}
\usepackage{color}
\usepackage{bm}
\usepackage{amssymb}
\usepackage{amsmath}
\usepackage{amstext}
\usepackage{amsthm}
\usepackage{amsfonts}
\usepackage{latexsym}

\newcommand{\ignore}[1]{}
\def\beq{\begin{equation}}
\def\eeq{\end{equation}}
\def\beqa{\begin{eqnarray}}
\def\eeqa{\end{eqnarray}}

\newcommand{\ket}[1]{\left| #1 \right>}
\newcommand{\bra}[1]{\left< #1 \right|}
\newcommand{\braket}[2]{\left< #1 \big |  #2\right>}

\begin{document}

\title{Topological phases of lattice bosons with a dynamical gauge field}
\author{David Ravent\'os}  
\affiliation{ICFO-Institut de Ciencies Fotoniques, The Barcelona Institute of Science
and Technology, 08860 Castelldefels (Barcelona), Spain}
\author{Tobias Gra{\ss}} 
\affiliation{ICFO-Institut de Ciencies Fotoniques, The Barcelona Institute of Science
and Technology, 08860 Castelldefels (Barcelona), Spain}
\author{Bruno Juli\'a-D\'iaz} 
\affiliation{Departament d'Estructura i Constituents de la Mat\`{e}ria,\\
Facultat de F\'{\i}sica, Universitat de Barcelona, 08028 Barcelona, Spain}
\affiliation{Institut de Ci\`encies del Cosmos, Universitat de Barcelona, IEEC-UB, Mart\'i i
Franqu\`es 1, 08028 Barcelona, Spain}
\affiliation{ICFO-Institut de Ciencies Fotoniques, The Barcelona Institute of Science
and Technology, 08860 Castelldefels (Barcelona), Spain}
\author{Luis Santos}
\affiliation{Institut f\"ur Theoretische Physik, 
Leibniz Universit\"at Hannover, Appelstr. 2, DE-30167 Hannover, Germany}
\author{Maciej Lewenstein}
\affiliation{ICFO-Institut de Ciencies Fotoniques, The Barcelona Institute of Science
and Technology, 08860 Castelldefels (Barcelona), Spain}
\affiliation{ICREA-Instituci\'o Catalana de Recerca i Estudis
Avan\c{c}ats, 08010 Barcelona, Spain}

\begin{abstract}

Optical lattices with a complex-valued tunnelling term have become 
a standard way of studying gauge-field physics with cold atoms. 
If the complex phase of the tunnelling is made density-dependent, 
such system features even a self-interacting or dynamical magnetic 
field. In this paper we study the scenario of a few bosons in 
either a static or a dynamical gauge field by means of exact 
diagonalization. The topological structures are identified computing 
their Chern number. Upon decreasing the atom-atom contact interaction, 
the effect of the dynamical gauge field is enhanced, giving rise to 
a phase transition between two topologically non-trivial phases.

\end{abstract}

\maketitle

\section{Introduction}

The external motion of a particle can be coupled to the dynamics of 
internal degrees of freedom via a gauge potential. The simplest example 
of this mechanism is that of an electrically charged particle moving 
in the presence of a background magnetic field. The gauge field imprints 
a complex $U(1)$ phase onto the wave function of the particle. The 
synthetic implementation of this mechanism in cold atomic systems has 
been envisaged since the early days of quantum 
gases~\cite{cooperwilkin,jakschzoller,mueller,fleischhauer,osterloh,cooper-aip,dalibard}, 
and has been realized successfully during the last 
years~\cite{lin,spielmanPRL,spielman-sobec,sengstock12,spielman-peierls,bloch-gauge}. 
Current quantum simulation with artificial gauge potentials are exploring 
the variety of interesting physics related to background gauge fields: 
spin liquid phases~\cite{sengstockXY}, topological phases evidenced 
by non-zero Chern numbers~\cite{bloch_chern}, or quantum Hall phases with 
edge currents~\cite{Mancini25092015,Stuhl25092015}. A long-term goal is the simulation 
of quantum electromagnetism or chromodynamics, that is, of models where 
matter interacts with dynamical fields, as described in 
Refs.~\cite{zohar12,luca13,banerjee,zohar,luca}. An intermediate step might 
be the realization of simpler but nevertheless dynamical gauge fields, 
engineering an occupation number-dependent tunnelling 
term~\cite{edmonds13,greschner14,10.1038/ncomms1353,om14,PhysRevLett.115.053002,PhysRevB.92.115120,om15,bermudez15}.

In this article, we consider a specific dynamical gauge field and apply 
exact diagonalization techniques to shed light on the involved interplay 
between the atoms' external degree of freedom and the system's $U(1)$ 
gauge potential. The atoms are confined to a two-dimensional optical 
lattice, where a gauge field is present due to a density-dependent 
complex phase of the tunnelling parameter $t$. Deep in the Mott phase, 
where density fluctuations are strongly suppressed, the gauge potential 
is static. We follow the system's evolution upon decreasing the ratio 
$U/t$, where $U$ parametrizes the strength of the repulsive on-site 
interactions. For sufficiently weak interactions, topological transitions, 
not present in the system with a static gauge field, are found in the 
system with a dynamical gauge potential. 

In our study the system is assumed to be close to filling one, where 
for large enough atom-atom interaction the Mott insulating state provides 
a vacuum-like configuration. In the strongly interacting regime, an 
extra-particle on top of the Mott insulator can be viewed as a single 
particle in a static gauge potential with a fixed magnetic flux per 
plaquette. This configuration therefore reproduces Hofstadter 
physics~\cite{hofstadter}. Due to computational limitations, our study 
addresses a 3$\times$3 lattice with $4\pi/3$ flux per plaquette. 
Twisted periodic boundary conditions allow for reducing finite-size 
effects. The low-energy subspace is clearly divided into three gapped 
bands. Chern number calculations demonstrate the non-trivial topological 
nature of the bands.
Since a hole in the Mott insulator does not feel any 
gauge potential, the extra-particle configuration also captures the 
behavior in a larger Mott insulator with a particle-hole excitation. Upon 
decreasing the interaction, we find deviations from this single-particle 
picture. For a dynamical gauge potential we find that the ground state 
undergoes a topological phase transition before it becomes topologically 
trivial in the limit $U \rightarrow 0$.

The article is organized as follows. First, in Sec:~\ref{Sec:Sys}, we 
describe our theoretical tools, including the density dependent Hamiltonian 
we are considering. 
Then in Sec.~\ref{sec:egaps} we present results for the different band gaps found, 
comparing the case of a dynamical field and the one of an static external 
field. The characterization of the topological properties by means of 
Chern numbers is presented in Sec.~\ref{sec:top}. 
In Sec.~\ref{sec:MFdiag} a phase diagram trough a 
Mean Field approach is presented in order to give an
intuitive idea of the behaviour of the system in the infinite size case.
Finally, in Sec.~\ref{sec:sum} we provide a brief summary and conclusions. 
In addition, the Appendix~\ref{Sec:Chern} includes the
procedure used to compute Chern numbers for 
the many-body bands to characterize the topological phases.

\section{Theoretical model}
\label{Sec:Sys}

\begin{figure}[t]
\resizebox{\columnwidth}{!}{\includegraphics{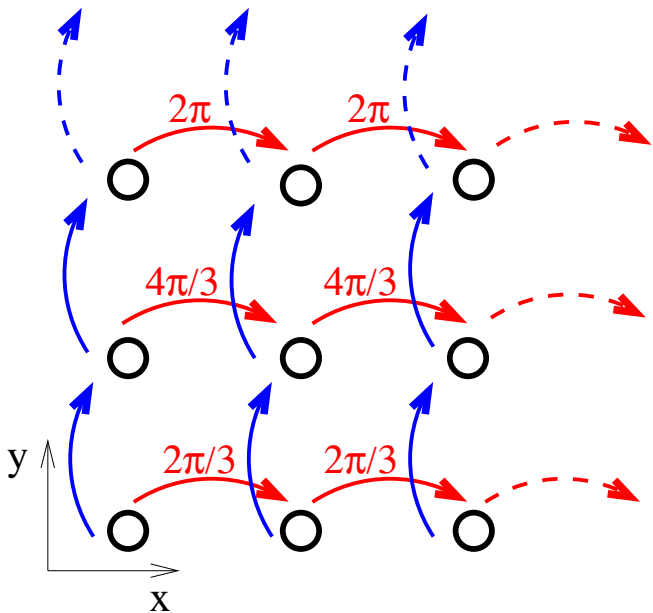}}
\caption{Brief rendition of the considered density dependent Hamiltonian. 
As an example we provide the phase acquired for the single particle case, 
$\hat{\cal H}_{\rm Landau}$, with $\varphi=2\pi/3$. The solid lines represent 
the tunnelling terms, the dashed ones correspond to the periodic boundary 
conditions considered. 
\label{fig:scheme}}
\end{figure}

Cold atoms in optical lattices are well described by a Hubbard model 
combining nearest-neighbour hopping processes and on-site interactions~\cite{mlbook}. 
The effect of a (synthetic) magnetic field is taken into account by a 
Peierls phase in the hopping parameter. For instance, if $\hat{b}_{k,l}$ 
($\hat{b}_{k,l}^\dagger$) denotes the annihilation (creation) of a particle 
at site $(k,l)$, the hopping term in a constant magnetic field 
with magnetic flux $\varphi$ per plaquette is written in the Landau gauge as
\begin{align}
\label{landau}
\hat{\mathcal{H}}_{\rm Landau} = 
- t \sum_{k,l} \left( { \rm e}^{{\rm i} \varphi l} \hat{b}^{\dagger}_{k,l} 
\hat{b}_{k+1,l} +\hat{b}^{\dagger}_{k,l} \hat{b}_{k,l+1} + {\rm h.c.} \right) \,.
\end{align}
Here, $t$ is a real-valued parameter associated 
with the kinetic energy of the particles. We consider a two-dimensional 
system of scalar bosons. Important quantities like the energy spectrum 
of the Hamiltonian are gauge-independent, that is, alternative hopping 
Hamiltonians with complex phases along the $x$-direction or along both 
the $x$- and $y$-direction would lead to the same results as long as the 
flux per plaquette remains the same. A schematic representation of the hopping 
structure is given in Fig.~\ref{fig:scheme}.

A possible implementation of Hamiltonians like $\hat{\mathcal{H}}_{\rm Landau}$ 
goes back to Ref.~\cite{jakschzoller}. In this paper, we are interested in 
a situation where the gauge field becomes dynamical, that is, the complex 
phase factor should in some form depend on the positions of the atoms. 
A simple dynamical gauge field is obtained by letting the phase depend 
on the occupation numbers, 
\begin{align}
\label{dyn}
 \hat{\mathcal{H}}_{\rm dyn}=&
- t \sum_{k,l} \Big( \hat{b}^{\dagger}_{k,l} 
{ \rm e}^{{\rm i} \varphi l \left(\hat{n}_{k,l}+\hat{n}_{k+1,l}\right)}\hat{b}_{k+1,l} \nonumber \\ 
& +\hat{b}^{\dagger}_{k,l} \hat{b}_{k,l+1}+  {\rm h.c.} \Big) \,.
\end{align}
The experimental implementation of density-dependent gauge fields as those of Hamiltonian (2)
can be done using similar techniques as those recently discussed in 
Refs.~\cite{PhysRevLett.115.053002,PhysRevB.92.115120}.
Particular details of how to implement 
it fall beyond the scope of the present article. 

This choice of the density dependent field is particularly attractive 
as it has one specific limit in which the topological properties of the 
system can be easily understood. Deep in the Mott insulating phase, where 
the number operators $\hat{n}_{k,l}$ can be replaced by an integer number 
$n$, this Hamiltonian reduces to the form of a $\hat{\mathcal{H}}_{\rm Landau}$. The amount 
of particle number fluctuations and thereby the dynamical features of 
the gauge potential are controlled by the interaction term, 
$\hat{\mathcal{H}}_{\rm int} = \frac{U}{2} \sum_{k,l} \hat{n}_{k,l} (\hat{n}_{k,l}-1)$. 
With this, the full Hamiltonian reads
\begin{align}
\label{eq:ham}
\hat{\mathcal{H}}= \hat{\mathcal{H}}_{\rm dyn}+ \hat{\mathcal{H}}_{\rm int} \,.
\end{align}
We will take an additional constraint on the Hilbert space, stemming from the 
implementation scheme described in Ref.~\cite{PhysRevLett.115.053002}, namely, the 
maximum occupancy per site will be set to two bosons.

To clarify our discussion we will compare our results to those obtained with an 
static field, that is, 
\beq
\hat{\mathcal{H}}_{\rm st} =  \hat{\mathcal{H}}_{\rm Landau}+ \hat{\mathcal{H}}_{\rm int} \,.
\eeq

\section{Energy gaps}
\label{sec:egaps}

\begin{figure*}[t]
\resizebox{2\columnwidth}{!}{\includegraphics{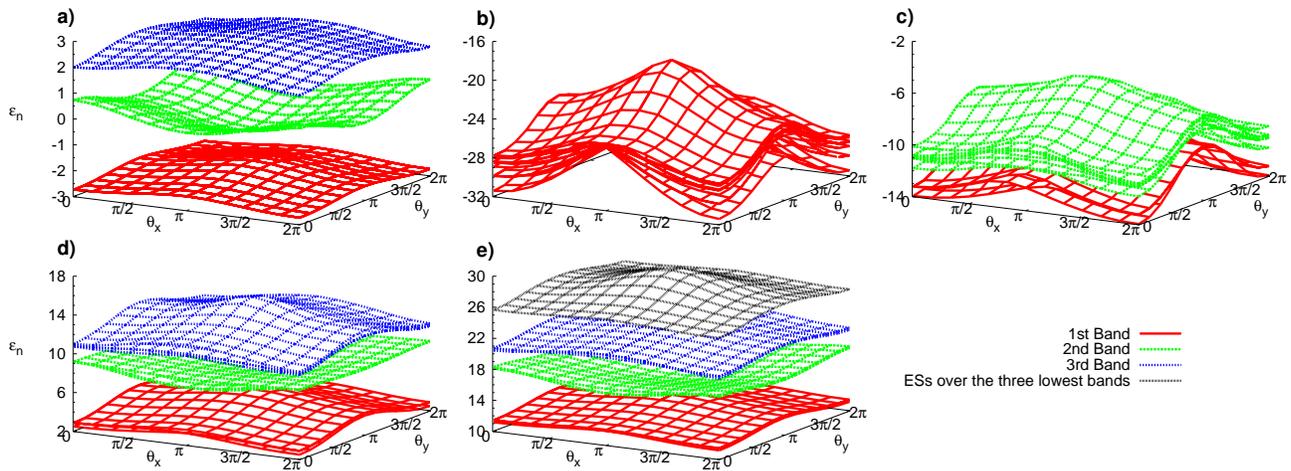}}
\caption{Energy spectrum as a function of the twisted 
boundary conditions for several systems under $\hat{\cal H}_{\rm st}$
with a flux per plaquette $\varphi = 2\pi/3$ in a $3 \times 3$ lattice.
Degenerated states forming bands have the same color.
a) Single particle case.
b)--e) Energies of the ten lowest eigenstates of the system with 10 particles for 
the interaction values: b) $U=0 t$, c) $U=4 t$, d) $U=13 t$ 
and e) $U=20 t$.
ESs means Excited States.
\label{fig:spectra_TwBC}}
\end{figure*}

We have concentrated on the filling case around one
by means of exact diagonalization. We have focused 
on a 3$\times$3 lattice at $\varphi=4\pi/3$, and take the interaction 
strength $U$ (in units of $t$) as the main tuning parameter. As 
argued above, this also controls the influence of the dynamical gauge 
field. To gain meaningful results despite the small system size, we apply 
twisted boundary conditions with twist angles $\theta_x$ and $\theta_y$. 
With this, the energy spectrum $\epsilon_i$ of the Hamiltonian becomes a 
function of the twist angles, $\epsilon_i(\theta_x,\theta_y)$. Degeneracies 
of different levels which would be lifted due to the finite system size 
manifest themselves in crossings of bands $\epsilon_i(\theta_x,\theta_y)$. 
Accordingly, we define the gap above a level $\epsilon_i$ as
\begin{equation}
\Delta \epsilon_i = 
{\rm min} \left[ \epsilon_{i+1} \left( \theta_x, \theta_y \right)
-\epsilon_{i} \left( \theta_x, \theta_y \right) \right] \,.
\end{equation}

\begin{figure}[t]
\resizebox{\columnwidth}{!}{\includegraphics{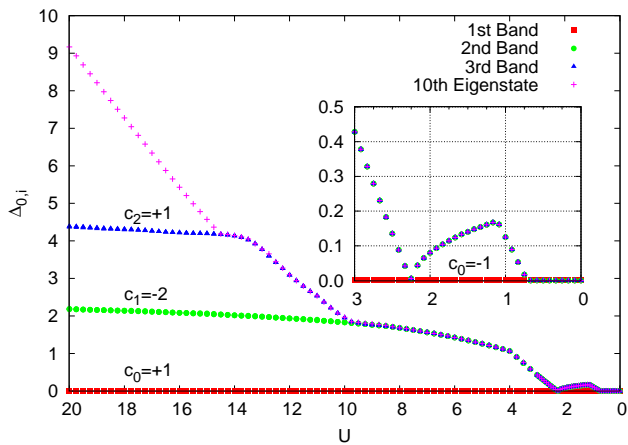}}
\caption{Gap above each band, [as defined in Eq.~(\ref{gap})], between consecutive 
eigenstates of $\hat{\mathcal{H}}$ as function of the on-site interaction 
parameter $U$.
The system is a $3\times3$ lattice with $10$ particles with the parameter 
$\varphi=2\pi/3$ and $t=1$. We take into account the 
Hilbert constraint to a maximum of two bosons per site.
The three lowest bands have degeneracy 3.
and the labels correspond to the Chern number of each band.
\label{fig:MinGapBand_S}}
\end{figure}

If $\Delta \epsilon_i$ is zero, that is, if band $i$ and band $i+1$ have 
(at least) one crossing, we consider these levels a degenerate manifold. 
To check whether the manifold is separated from higher levels by a 
gap, we then have to consider $\Delta \epsilon_{i+1}$. In general, 
the gap above a $k$-fold manifold including the levels $i,\dots,i+k$ 
is defined as
\begin{align}
\label{gap}
\Delta_{i,i+k} =  \sum_{j=i}^{i+k-1} \Delta \epsilon_j  \,.
\end{align}

\begin{figure}[t]
\resizebox{\columnwidth}{!}{\includegraphics{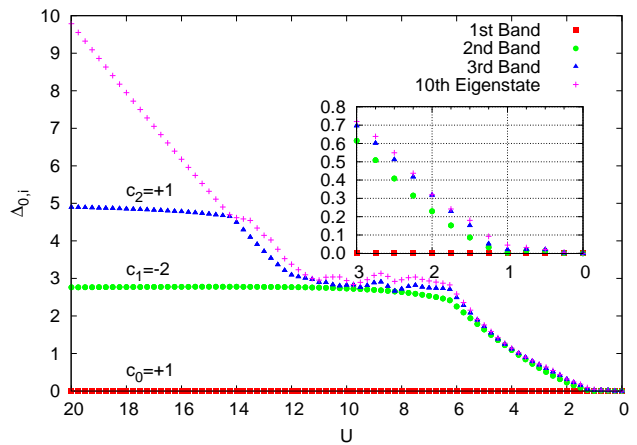}}
\caption{Same description as in Fig.~\ref{fig:MinGapBand_S}, but with 
the dynamical gauge field replaced by an external magnetic field with a 
flux per plaquette $\varphi=4\pi/3$. That is,
$\hat{\cal{H}}\left( \phi\right) \rightarrow \hat{\cal{H}}_{\rm st}\left( 2\phi\right)$.
\label{fig:MinGapBand_E}}
\end{figure}

\subsection{Case of one excess particle}

\begin{figure*}[t]
\includegraphics[width=0.4\textwidth,angle=-90]{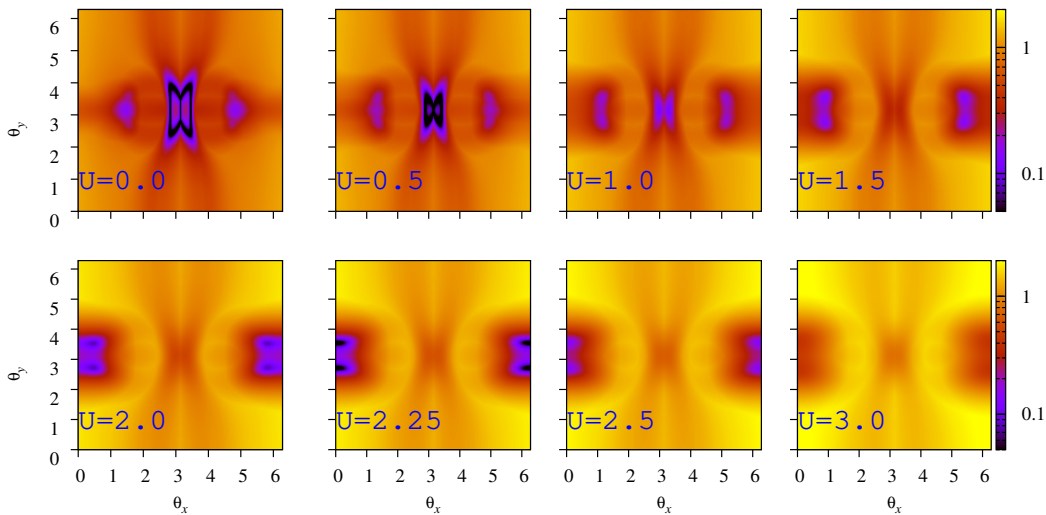}
\caption{Energy difference between the third and fourth states in the 
spectrum of a system under $\hat{\cal H}$ with 10 particles in a $3 \times 3$ lattice, signaling the gap between the ground state manifold and next
excited state. The different panels correspond to different values of 
$U$. In all cases, $\varphi=2\pi/3$ and $t=1$.
\label{fig:gap}}
\end{figure*}

We start our analysis with the tunnelling of a system with one particle more than the number of sites.
That is, in our $3 \times 3$ lattice we consider $N=10$ bosons.
On the strongly interacting side, this is equivalent to having a single 
particle on top of a fluctuating vacuum. For large $U$, fluctuations are 
strongly suppressed, and the kinetic Hamiltonian (\ref{dyn}) reduces to 
the one of a particle in a static magnetic field, Eq.~(\ref{landau}), 
with flux 2$\varphi$. Accordingly, the physics of a single particle in 
a magnetic field should describe the low-energy behavior of our system. 
Indeed, no difference is seen between the shape of the single-particle spectrum of 
the Hamiltonian (\ref{landau}), Fig.~\ref{fig:spectra_TwBC} a), 
and the (low-energy part) of the one of the many-body spectrum of Hamiltonian 
(\ref{eq:ham}) at large $U$, Fig.~\ref{fig:spectra_TwBC} e).
In both cases, we find the energy spectrum to be split into three gapped manifolds, 
each of them consisting of three states. In the many-body system, a gapless 
high-energy manifold lies above the third band.

Deviations from this structure appear when $U$ is decreased, see 
Fig.~\ref{fig:spectra_TwBC}(b--e) and Fig.~\ref{fig:MinGapBand_S}. 
The dynamical mechanism is the following. As  $U$ is decreased, the 
number of holon-doublon excitations increases, the single-particle picture 
described above is no-longer valid. First, for $U \approx 15$, the gap 
between the third band and the high-energy manifold closes, as the first 
doublon-holon excitations have the same energy as the third single particle 
state. Subsequently, at $U\approx 10$, also the gap to the second band is closed. 
These gap closings indicate phase transitions in excited states. 
At  $U\simeq 2.25$, also the gap to the lowest band is closed. Thus, 
up to $U\simeq 2.25$ the ground state manifold has a topological structure 
similar to the case of a single particle subjected to an external magnetic field 
of $2 \varphi$. This value of $U$ is a bit higher than the value at which 
we found a gappless phase for the filling one case described below. Thus, 
the main picture of a single particle on top of a Mott insulating background 
is consistent.

Remarkably, further lowering the value of the interaction another 
threefold-degenerate gapped manifold appears for $0.67 \lesssim U \lesssim 2.25$. 
Only for $U\lesssim0.67$ the system enters in a gapless phase. We note that 
for $0.67 \lesssim U \lesssim 2.25$ the gap is small, of the order of 10\% of 
the involved energy scales. It is a merit of the twisted boundary conditions 
that the three lowest states are clearly identified as an adiabatically 
connected manifold, separated from the other levels by a gap. In fact, if 
we look at the system for a fixed value of $\theta_x$ and $\theta_y$, or 
alternatively for open boundary conditions, the gap cannot be distinguished 
from the energy splitting between states in the degenerate manifold. The 
evolution of the gap between the ground state manifold and the next excited 
state for the all values of $\theta_x$ and $\theta_y$ is given in Fig.~\ref{fig:gap}.
The gap above a manifold as function of $U$ is shown in Fig.~\ref{fig:MinGapBand_S} .
For $U\simeq 0.67$ the gap closes at $(\theta_x,\theta_y)\simeq (\pi,\pi)$. The next 
closing, for $U\simeq 2.25$ appears close to $(\theta_x,\theta_y)\simeq (0,\pi)$. 
This could diminish the prospects for an experimental detection of this 
phase in the plane geometry, but since an experiment would realize a much 
bigger system, there is hope that finite-size degeneracy splitting would 
be sufficiently small to identify the finite gap.

In Fig.~\ref{fig:MinGapBand_E}, we contrast our findings to the scenario 
with static magnetic field. As expected, at large $U$ the differences between 
Fig.~\ref{fig:MinGapBand_S} and~\ref{fig:MinGapBand_E} are minor. Also for a static
magnetic field, increasing $U$ subsequently closes the gaps above the third 
and the second band. However, the gap above the ground state remains finite 
up to $U \simeq 1$ and, for $U<1$, it vanishes.

\begin{figure}[t]
\resizebox{\columnwidth}{!}{\includegraphics{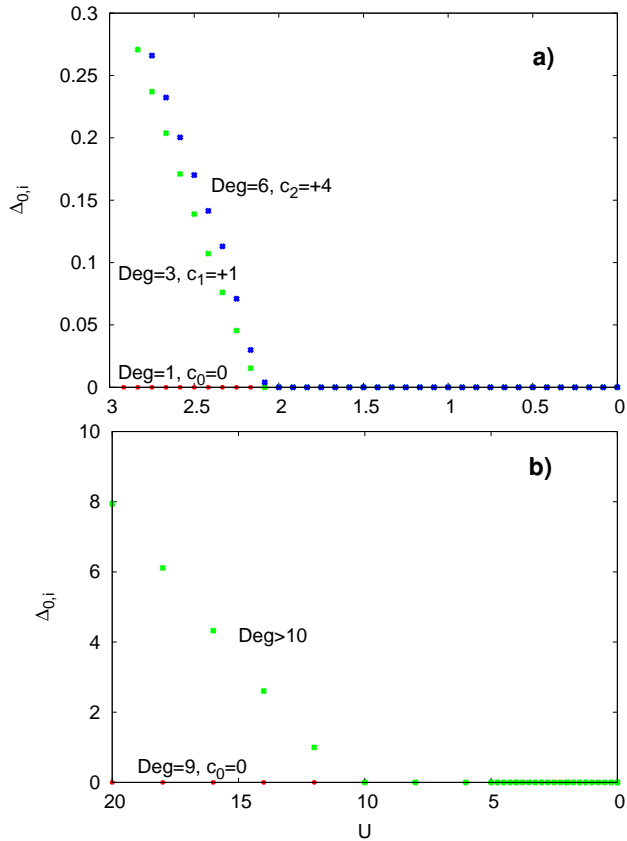}}
\caption{Energy gap, see Eq.~(\ref{gap}), between consecutive eigenstates 
of $\hat{\mathcal{H}}$ as function of the on-site interaction parameter 
$U$. We take into account the Hilbert constraint to a maximum of two bosons 
per site. The upper and lower panels are for $N=9$ and $N=8$, respectively. 
\label{fig:n89}}
\end{figure}

\subsection{Mott insulator}

At precisely filling one, for 9 particles on 9 lattice sites, see the upper panel of Fig.~\ref{fig:n89}, 
we find a unique gapped ground state for $U \gtrsim 2.1$, which is connected 
to the Mott insulator as an exact solution for $U\rightarrow \infty$. This 
phase is trivial in the sense that it corresponds to a vacuum, where 
deviations from integer filling exist only as fluctuations. For $U\lesssim 2.1$, 
we find a gapless phase, that is, despite the presence of the dynamical 
gauge field, no topological structure protected by an energy gap emerges 
in this scenario.

The first and second excited bands are three- and 
six-fold degenerate, respectively. 
They are topologically non-trivial and their Chern numbers are $+1$ and $+4$.
This excited bands coincide, in degeneracy and topology, 
with the lowest band of the non-interacting systems with one and two particles in the same lattice,
as they are explained in Section~\ref{sec:top}.~A.
These excited bands can be understood as one and two particle-hole excitations 
on the top of the Mott insulator,
when the particle feels an effective static magnetic field and the hole do not.

\subsection{Case of one hole}

We also study the tunnelling of a single hole. That is, 
in our $3 \times 3$ lattice we consider $N=8$ bosons. The gap structure we 
find is shown in Fig.~\ref{fig:n89}~b). As expected, we find that increasing 
the interaction up to $U\simeq 10 t$, a gap opens between the 
nine-fold degenerate manifold, understood as one hole moving in the 
Mott-insulating background, and the rest of the states. The ground 
state manifold is found to have a trivial topological order.

\section{Topological phases}
\label{sec:top}

In the previous section we have discussed the energy gaps 
appearing for the case of one excess particle, the 
filling one, and the one hole case. The only case in which we have found 
non-trivial topological structures is for the case of one excess 
particle. In the following we present the Chern number obtained compared 
to the case of an external field of flux $4\pi/3$. 

\subsection{Single particle and non-interacting cases}

First, we calculate the Chern numbers of the single-particle system 
described by $\hat{\mathcal H}_{\rm Landau}$, that is, of the bands shown 
in Fig.~\ref{fig:spectra_TwBC} a). We obtain the values $\{1,-2,1\}$. 
In this case, the calculation can either be done via Fourier transformation, 
taking the parameters $k_1$ and $k_2$ to be components of the wave 
vector~\cite{TKNN}, or with twisted boundary conditions, taking the 
twist angles $\theta_x$ and $\theta_y$ as parameters $k_1$ and $k_2$~\cite{niu}. 
In the latter case, the discretization of parameter space is 
arbitrary, but we observe quick convergence of the Chern numbers 
to fixed numbers upon refining the discretization.

The non-interacting case can be related to the single particle case although 
some caution should be exercised. For instance, direct computation of the 
Chern number of the ground state manifold for $N=2$, $3$, and $4$ particles in the $3 \times 3$ 
lattice we consider gives $c=4, 10$ and $20$, respectively. These can be 
obtained by noting that due to the bosonic symmetry, we have a combinatorial 
factor stemming from the number of times the Fock basis covers the threefold 
degenerate band. This can be evaluated giving, 
\begin{equation}
c_{0}^{\left(N\right)}= \frac{N}{3} \binom{N+3-1}{ N} c_{0}^{\left(1\right)}=c_{0}^{\left(1\right)} \, \binom{N+2}{3} \,,
\end{equation}
where $c_{0}^{\left(1\right)}$ is the single particle Chern number of the GS manifold, $c_{0}^{\left(1\right)}=1$. 

\begin{figure}[t]
\resizebox{\columnwidth}{!}{\includegraphics{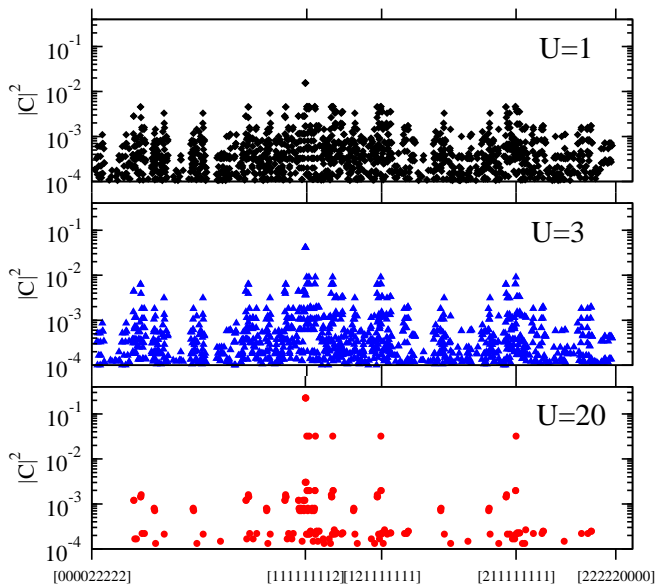}}
\caption{Squared coefficients of the ground state of the density dependent 
${\cal \hat{H}}$ for $\theta_x=\theta_y=0$ in the Fock basis in lexicographical order. 
A few notable states, one particle on top of a Mott insulator, are marked. The three 
panels corresponds to three values of the interaction, $U=1, 3$ and $20$. For these 
values, the Chern number of the ground state manifold is $-1$, $+1$ and, $+1$, respectively.
\label{fig:states}}
\end{figure}

\subsection{Interacting many-body case}

To calculate the Chern numbers of many-body states, we exclusively resort 
to the twisted boundary conditions. For the three gapped manifolds appearing 
at $U \lesssim 15$, see Fig.~\ref{fig:MinGapBand_S}, we obtain the same 
Chern numbers as for the single particle bands: $\{1,-2,1\}$. These numbers 
remain constant for each manifold until the closing of the corresponding gap. 
Upon closing the gap, the second and the third bands simply merge with the 
energy continuum, for which no Chern number can be computed. This is easily 
understood as for large enough interactions the many-body ground state is 
well described as consisting on a Mott-insulating background plus one particle. 
The lower band is given by the energy of the extra particle in the presence 
of an external field with flux $4 \pi/3$. The closing of the bands in the 
higher part of the spectrum comes from the first particle-hole excitations 
which eventually degenerate with excitations of the excess particle. 
Note that even though this simple picture provides a compelling explanation it 
is quite remarkable that albeit the many-body state changes, as shown in 
Fig.~\ref{fig:states}, as $U$ is decreased, 
the topology of the band does not change for a broad range of $U$. 

In contrast with the above, the gap closing of the ground state at 
$U\approx 2.25$ separates two gapped regions, see in particular the inset 
of Fig.~\ref{fig:MinGapBand_S}. Interestingly, we find that upon closing 
the band gap, the ground state Chern number changes its sign from $1$ to 
$-1$. This demonstrates that a topological phase transition between two 
distinct, but topologically non-trivial phases is taking place. The second 
gap closing, at $U\approx 0.67$, merges the ground state manifold with the 
energy continuum which, in this sense, is a transition to a topologically 
trivial (gapless) phase.

\subsection{Static field case}

Finally, we note also that the three gapped manifolds found for a system 
with static magnetic field, with $\phi=4 \pi/3$, are characterized by the 
same Chern numbers $\{1,-2,1\}$, without any transitions to distinct gapped 
phases. As seen in Fig.~\ref{fig:MinGapBand_E}, the arguments exposed 
above also apply to this case and the picture of a single particle 
on top of a Mott-insulator is perfectly valid. The only relevant 
difference appears for low interaction energies. In this case, the 
Mott insulating phase seems to survive down to lower values of the 
interaction as compared to the density dependent case. Thus, density 
dependence phases favor the existence of superfluid regimes at larger 
interactions than in the static case. Also we find no trace of the 
first excitation being a topological phase with $c_1=+1$ in the region 
$2.25 \gtrsim U \gtrsim 0.67$. 
In this case the limit $U=0$ can be understood from the single-particle 
calculation: The ground states for $N$ bosons are just arbitrary 
distributions on the $M=N_{\rm s}/q$ states belonging to the lowest 
energy band in a lattice with $N_{\rm s}$ sites at magnetic flux $2\pi/q$. 
This leads to a macroscopic ground state degeneracy (of 63 states in 
our case with $N=10$, $N_{\rm s}=9$, and $q=3$), for which no meaningful 
Chern number can be defined. Recent ``Chern number'' measurements in 
non-interacting bosonic quantum gases~\cite{bloch_chern} consider 
the Hall drift for unique but gapless many-body states, and define 
as a ``Chern number'' the average over different states.

\section{Mean field Phase diagram}
\label{sec:MFdiag}

In order to get a picture of the phase diagram, we have adapted the Mean-field calculation of Ref.~\cite{10.1038/ncomms1353}
to the Hamiltonians of interest. At first, we include a chemical potential term $-\mu \sum_{i,j} \hat{n}_{i,j}$.
With the convenient substitutions $\hat{c}_{i,j}\equiv e^{{\rm i} \phi j \hat{n}_{i,j}} \hat{b}_{i,j}$ and
$\hat{d}_{i,j}\equiv e^{-{\rm i} \phi j \hat{n}_{i,j}} \hat{b}_{i,j}$, the Hamiltonian in Eq.~(\ref{eq:ham}) looks like,

\begin{align}
\label{eq:hamsubs}
 \hat{\mathcal{H}}=&
\sum_{k,l} \Big\{ - t \Big( \hat{d}^{\dagger}_{k,l} \hat{c}_{k+1,l} +\hat{b}^{\dagger}_{k,l} \hat{b}_{k,l+1} +  {\rm h.c.}  \Big)\nonumber \\ 
& +\hat{n}_{k,l}\left[\left(\hat{n}_{k,l} -1 \right) \frac{U}{2} -\mu\right]  \Big\} \,.
\end{align}

At $t=0$, all the sites are independent and the GS can exactly be represented with a Gutzwiller ansatz,
\begin{equation}
| \Psi_0 \rangle = \bigotimes_{k,l}^{N_{\rm s}} | \psi \rangle_{k,l},
\qquad
| \psi \rangle_{k,l}  = \sum_{m=0}^{\infty} f_{k,l}^{\left(m \right)} | m \rangle_{k,l},
\end{equation}
where $m$ is the number of particles in a site.
Then, the energy due to each site filled with $m$ particles is $\epsilon_{m} = U \left[ \frac{1}{2}\left(m-1 \right) -\frac{\mu}{U}\right]m$.
and the energy of adding and subtracting one boson is,
\begin{equation}
\epsilon_{m+1} - \epsilon_{m} = U \left( m-\frac{\mu}{U} \right),
\qquad
\epsilon_{m-1} - \epsilon_{m} = U \left( \frac{\mu}{U} -  m+1 \right),
\end{equation}
respectively.

The MF is obtained by decoupling the hopping terms as 
$\hat{d}_{i,j}^{\dagger} \hat{c}_{i+1,j}\approx \alpha_{3,j}^{*} \hat{c}_{i+1,j} + \alpha_{2,j} \hat{d}_{i,j}^{\dagger} - \alpha_{3,j}^{*} \alpha_{2,j}$
and $\hat{b}_{i,j}^{\dagger} \hat{b}_{i,j+1} \approx \alpha_{1,j}^{*} \hat{b}_{i,j+1} + \alpha_{1,j+1} \hat{b}_{i,j}^{\dagger} - \alpha_{1,j}^{*} \alpha_{1,j+1}$,
with the order parameters $\alpha_{1,j}\equiv \langle \hat{b}_{i,j} \rangle$,  $\alpha_{2,j}\equiv \langle \hat{c}_{i,j} \rangle$ and, $\alpha_{3,j}\equiv \langle \hat{d}_{i,j} \rangle$.
Then, the Hamiltonian in Eq.~(\ref{eq:hamsubs}) becomes,
\begin{align}
\label{eq:hamMF}
 \hat{\mathcal{H}}=&
 -N_{\rm x} t \sum_j \left( \alpha_{3,j}^{*} \alpha_{2,j} +\alpha_{1,j}^{*} \alpha_{1,j+1} +{\rm h.c.}\right)
+\sum_{k,j} \hat{h}_{k,j},
\end{align}
with the local Hamiltonian
\begin{align}
\hat{h}_{k,j} \equiv  \hat{n}_{k,j} \left[U\left(\hat{n}_{k,j} -1 \right)/2 -\mu\right] - t \hat{T}_{k,j},
\end{align}
where 
$
\hat{T}_{k,j} \equiv \alpha_{3,j}^{*} \hat{c}_{k,j} +\alpha_{2,j} \hat{d}^{\dagger}_{k,j} + \alpha_{1,j-1}^{*} \hat{b}_{k,j} +\alpha_{1,j+1} \hat{b}^{\dagger}_{k,j} +  {\rm h.c.}
$
and $N_{\rm x}$ is the size of the system in the $x$-direction.
The Hamiltonian $\hat{h}_{k,j}$ has a trivial solution when $\alpha_{\gamma,j} = 0$, $\gamma = 1,2,3$ since the particle number fluctuations vanish at the Mott insulating phase.

When the kinetic term is negligible ($t \ll U$), the entire system is described with the basis of states with $m$ particles per each site $\left( k,j \right)$, $| m \rangle$.
The GS is determined by $\mu$: it is the local state $|m\rangle$ when $m-1<\mu<m$.
Since we want to draw the Mott lobes, we include the single Fock state and particle-hole excitations in that region of the diagram.
Then, since we search the boundaries close to the trivial solution, $\left|\alpha_{\gamma,j}\right| \ll 1$, and the kinetic term can be treated perturbatively.
Up to first perturbation order, the local wavefunction $| \Psi \rangle$ can be written as $| \psi ^{\left( 0 \right)}\rangle+| \psi ^{\left( 1 \right)}\rangle$,
being $| \psi ^{\left( 0 \right)}\rangle = | m \rangle$ and
\begin{widetext}
\begin{align}
| \psi ^{\left( 1 \right)}\rangle = & -t \sum_{m^{\prime}} \frac{\langle m^{\prime}|\hat{T}_{k,j}
| m\rangle}{\epsilon_{m^{\prime}}-\epsilon_{m}} | m^{\prime}\rangle \nonumber \\
= & \frac{t}{U}\frac{\sqrt{m}\left[ \alpha_{3,j}^{*} e^{{\rm i} \phi j \left(m-1 \right)} + \alpha_{2,j}^{*} e^{-{\rm i} \phi j \left(m-1 \right)} +
 \alpha_{1,j-1}^{*}+\alpha_{1,j+1}^{*}\right]}{\frac{\mu}{U}-\left(m-1\right)}| m-1\rangle \nonumber \\
 &+\frac{t}{U}\frac{\sqrt{m+1}\left[ \alpha_{3,j} e^{-{\rm i} \phi j m} + \alpha_{2,j} e^{{\rm i} \phi j m} +
 \alpha_{1,j-1}+\alpha_{1,j+1}\right]}{m -\frac{\mu}{U}}| m+1\rangle
\end{align}
\end{widetext}

The first order perturbation about the solution $\alpha_{\gamma, j}=0$ is convenient here, since 
the self-consistency equations define a linear map $\alpha_{\gamma, j} = \Lambda_{\gamma, j}^{\gamma^{\prime}, j^{\prime}} \alpha_{\gamma^{\prime}, j^{\prime}}$.
Then, when the largest eigenvalue of $\Lambda$, $\lambda_0$, is larger than $1$, the trivial solution is no longer stable.
So, the boundary is found to be at $\lambda_0=1$.
The self-consistency relations $\alpha_{1,j}=\langle \Psi | \hat{b}_{k,j} |\Psi\rangle$, $\alpha_{2,j}=\langle \Psi | \hat{c}_{k,j} |\Psi\rangle$ and,
$\alpha_{3,j}=\langle \Psi | \hat{d}_{k,j} |\Psi\rangle$ give,

\begin{align}
&\alpha_{1,j} = \frac{t}{U} \left[A \left( \alpha_{1,j-1}+\alpha_{1,j+1} \right) + f_{j}\left(\phi\right) \alpha_{2,j} + f_{j}\left(-\phi\right) \alpha_{3,j} \right] \nonumber \\
&\alpha_{2,j} = \frac{t}{U} \left[f_{j}\left(\phi\right) \left( \alpha_{1,j-1}+\alpha_{1,j+1} \right) + f_{j}\left(2 \phi\right) \alpha_{2,j} + A \alpha_{3,j} \right] \nonumber \\
&\alpha_{3,j} = \frac{t}{U} \left[f_{j}\left(-\phi\right) \left( \alpha_{1,j-1}+\alpha_{1,j+1} \right) + A \alpha_{2,j} + f_{j}\left(-2\phi\right) \alpha_{3,j} \right]
\label{eq:selfcons}
\end{align}
with
\begin{align*}
f_{j}\left(\phi\right) \equiv \left[ A + B \left(e^{-{\rm i} \phi j}-1\right)\right] e^{-{\rm i} \phi j m} \nonumber \\ 
A \equiv \frac{\frac{\mu}{U} + 1 }{\left[\frac{\mu}{U}-\left( m-1 \right)\right]\left[m-\frac{\mu}{U} \right]},
\qquad
B \equiv \frac{m}{\frac{\mu}{U}- \left( m-1 \right) }.
\end{align*}
For the case of the static magnetic field, the corresponding function $f_{j} ^{\rm st}\left(\phi\right)$ reduces to $ A e^{-{\rm i} \phi j}$.

\begin{figure}[t]
\resizebox{\columnwidth}{!}{\includegraphics{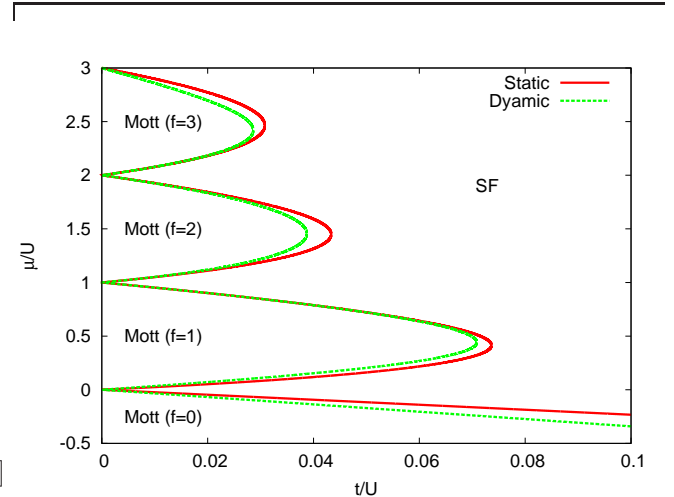}}
\caption{Phase boundary between the Mott insulator phase and the superfluid phase
for the external and dynamical magnetic fields according to the Gutzwiller ansatz 
with the MF approach in the 1st perturbation order in the hopping $t$.}
\label{fig:MF_Diags}
\end{figure}

The Fock space populations of the GS of the system, Fig.~\ref{fig:states}, have revealed its structure:
the Fock states which have a particle on top of a MI in the same row are equally populated.
Then, we have tried an ansatz which is  
translationally invariant along the $x$-direction and have a 3-unit cell in the $y$-direction.
So, in Eq.~(\ref{eq:selfcons}), $j=1,2,3$, without periodic boundary conditions. Then, those relations define a linear system of nine coupled linear equations,
being $\alpha_{\gamma,j}$ the variables.
Once the matrix of the system is diagonalized as function of $t$, for a numeric value of $U$, $\mu$ (and its corresponding integer $m$), 
the expression of $\lambda_0$ is set to $1$ and then the equation is solved for $t$.
Finally, the phase boundary is obtained as a collection of points $\left( \mu, t \right)$.

We find the Mott lobes, shaped as usual in the MF approach, see Fig.~\ref{fig:MF_Diags}. The values of the boundary do not correspond to the ones of the MF for the 2D lattice,
but they are closer to the ones of the 1D case, see Ref.~\cite{PhysRevA.79.013614}. The structure of the GS state has revealed this to be closely related to the fact 
that the magnetic fields are in the Landau gauge.
Our analysis also shows that the trial state is slightly more robust upon decreasing hopping $t/U$
in the dynamic field case than in the static one.
This finding qualitatively agrees with our results for the gap separation in the exact diagonalization analysis:
As seen in Figs.~\ref{fig:MinGapBand_S} and ~\ref{fig:MinGapBand_E}, the SF regime corresponds to the gapless phase at small $U$,
which extends to $U=0.8$ in the dynamic case, and $U=1.2$ in the static case.
For $\mu<0$, the boundaries of the dynamic field case and the 2D non-magnetic case coincide, as expected.

\section{Summary and conclusions}
\label{sec:sum}

We have studied topological properties of a bosonic quantum gas with 
an experimentally feasible, synthetic dynamical gauge field. The Mott 
insulating phase provides a trivial vacuum, above which we study 
the one-particle excitations, forming gapped energy bands. Decreasing 
the interactions, we first observe transitions in the excited bands, 
from topologically non-trivial phases to gapless phases. In this respect, 
the system behavior does not differ from the one of a system with static 
magnetic field. A particular feature of the dynamic gauge field is a 
topological transition in the ground state, in which the sign of the 
Chern number is inverted. 
The fact that in our proposal the length of the system in one dimension 
is very small could be accomplished in a experimental set-up using synthetic dimensions.

\textit{ Acknowledgements.}
We thank S. Greschner, T. Mishra, and M. Rizzi for discussions.
Financial support from EU grants EQuaM (FP7/2007-2013 Grant No. 323714), 
OSYRIS (ERC-2013-AdG Grant No. 339106), QUIC (H2020-FETPROACT-2014 No. 641122) and,
SIQS (FP7-ICT-2011-9 No. 600645), 
MINECO grant FOQUS (FIS2013-46768-P) and ``Severo Ochoa'' Programme (SEV-2015-0522),
Generalitat de Catalunya grants 2014 SGR 401 and 2014 SGR 874 and, Fundaci\'o Cellex is acknowledged.
B.J-D.\ is supported by the Ramon y Cajal program.
L.S. thanks the Center for Quantum Engineering and Space-Time Research, and the DFG Research Training Group 1729.

\bibliographystyle{apsrev4-1}

\appendix

\section{Evaluation of the Chern number}

\label{Sec:Chern}

The twisted boundary conditions are particularly useful to characterize 
topological phases. They allow one to define Chern numbers in an 
interacting many-body system~\cite{niu}. Quite generally, the Chern number 
is defined for the energy levels $n$ of a Hamiltonian 
${\cal H}(k_1,k_2)={\cal H}(k_1+2\pi,k_2)={\cal H}(k_1,k_2+2\pi)$, which 
periodically depends on two parameters $k_1$ and $k_2$ in the following way,
\begin{equation}
c_n=\frac{1}{2 \pi {\rm i}}\int_0^{2\pi} dk_1\int_0^{2\pi} dk_2 \; F_{12}^{\left(n\right)}\left(k_1,k_2\right)
\end{equation}
where the Berry connection ${\cal A}_{\mu}^{\left(n\right)}(k_1,k_2)$ ($\mu=1,2$) 
and the associated strength $F_{12}^{\left(n\right)}(k_1,k_2)$ are given by
\begin{eqnarray}
{\cal A}_{\mu}^{\left(n\right)}(k_1,k_2)=\bra{n(k_1,k_2)}
\partial_{\mu}\ket{n(k_1,k_2)}\\
F_{12}^{\left(n\right)}(k_1,k_2)=\partial_1 {\cal A}_{2}^{(n)}(k_1,k_2)
-\partial_2 {\cal A}_{1}^{\left(n\right)}(k_1,k_2)
\end{eqnarray}
with $\ket{n\left(k_1,k_2\right)}$ being the $n$th normalized eigenvector.

Following the method of Fukui {\it et al.}~\cite{2005JPSJ_74.1382M}, the Chern numbers 
can conveniently be calculated by discretizing the parameter space, 
\begin{equation}
\tilde{c}_n=
\frac{1}{2 \pi {\rm i}}\sum_{k_1} \sum_{k_2} \tilde{F}_{12}^{\left(n\right)}\left(k_1,k_2\right)
\label{chern}
\end{equation}
with the lattice field strength,
\begin{equation}
\begin{split}
\label{F}
& \tilde{F}_{12}^{\left(n\right)}\left(k_1,k_2\right)
=\ln{\left[\frac{U_1^{\left(n\right)}(k_1,k_2) \, U_2^{\left(n\right)}(k_1+dk_1,k_2)}
{U_1^{\left(n\right)}(k_1,k_2+dk_2) \, U_2^{\left(n\right)}(k_1,k_2)}\right]},\\
& -\pi < \frac{1}{\rm i} \tilde{F}_{12}^{\left(n\right)}\left(k_1,k_2\right) \le \pi
\end{split}
\end{equation}
being $dk_{\mu}$ the resolution of each parameter and $U_{\mu}^{\left(n\right)}$ the link 
variables from the eigenstates of the $n$th band,
\begin{equation}
\label{U}
U_{\mu}^{\left(n\right)} \equiv 
\frac{\braket{n(k_1,k_2)} {n(k_1+dk_1\delta_{1,\mu},k_2+dk_2\delta_{2,\mu})}}
{\left|\braket{n(k_1,k_2)} {n(k_1+dk_1\delta_{1,\mu},k_2+dk_2\delta_{2,\mu})}\right|}\,.
\end{equation}

A special case which is important for our purposes concerns the Chern 
number of degenerate bands. Since the eigenstates are not unique in 
the degenerate points, we cannot associate Chern numbers to individual 
states. For $M$ degenerate or quasi-degenerate states, we consider 
the multiplet $\psi = \left(\ket{n_1} \cdots \ket{n_M} \right)$ to define 
a non-Abelian Berry connection ${\cal A}=\psi^{\dagger}d\psi$, which is an 
$M \times M$ matrix-valued one form associated to $\psi$. Then, we 
consider the overlap matrix
\begin{equation}
\left[u_{\mu}^{\left(n\right)}\right]_{ij} \equiv 
\braket{n_i(k_1,k_2)} {n_j(k_1+dk_1\delta_{1,\mu},k_2+dk_2\delta_{2,\mu})},
\end{equation}
in order to properly define the link variables
\begin{equation}
U_{\mu}^{\left(n\right)} \equiv 
 \frac{\det{ \left[u_{\mu}^{\left(n\right)}\right]}}
{\left|\det{ \left[u_{\mu}^{\left(n\right)}\right]}\right|}
\end{equation}
Finally, the Chern number $\tilde{c}_{\psi}$ and field strength are calculated using 
Eqs.~(\ref{chern}) and (\ref{F}).

\end{document}